\documentclass[10pt]{article}
\usepackage{graphicx}

\title{Maximal Independent Sets in Generalised Caterpillar Graphs}

\author{Neethi K.S.\thanks{Presently with Microsoft, India}
{\normalsize{and }} Sanjeev Saxena\thanks{E-mail: ssax@cse.iitk.ac.in}\\
Dept. of Computer Science and Engineering,\\ Indian Institute of
Technology,\\ Kanpur, INDIA-208 016}

\date{\today}

\begin{document}
\maketitle

\subsection*{\centering{Abstract}}

A caterpillar graph is a tree which on removal of all its pendant
vertices leaves a chordless path. The chordless path is called the
backbone of the graph.  The edges from the backbone to the pendant
vertices are called the hairs of the caterpillar graph. Ortiz and
Villanueva (C.Ortiz and M.Villanueva, Discrete Applied Mathematics,
160(3): 259-266, 2012) describe an algorithm, linear in the size of the
output, for finding a family of maximal independent sets in a caterpillar
graph.

In this paper, we propose an algorithm, again linear in the output size,
for a generalised caterpillar graph, where at each vertex of the
backbone, there can be any number of hairs of length one and at most one
hair of length two.

{\textbf{Keywords:}} Maximal Independent Set; MIS; Caterpillar Graphs;
Generalised Caterpillar Graphs; Generating all MIS; Algorithm 

\section{Introduction}

A caterpillar graph $C(P_k)$ (See Figure 1) is a tree which on removal of
all its pendant vertices (vertices $h_i$ and $l_j$ in the figure) results
in a chordless path $P_k=\{v_1,v_2,..., v_k\}$ of $k$ vertices. The path
$P_k$ is called the backbone of the caterpillar graph $C(P_k)$ and the
edges from the backbone to the pendant vertices (edges $(v_i,h_i)$ and
$(v_j,l_j)$ in Figure 1) are called its hairs.  In $C(P_k)$ all hairs are
of length one. Harary and Schwenk [3], introduced Caterpillar graphs by
saying:\\

``Caterpillar is a tree which metamorphoses into a path when its
cocoon of endpoints is removed''.

\begin{figure}[h]
\centering
\includegraphics[width=2in]{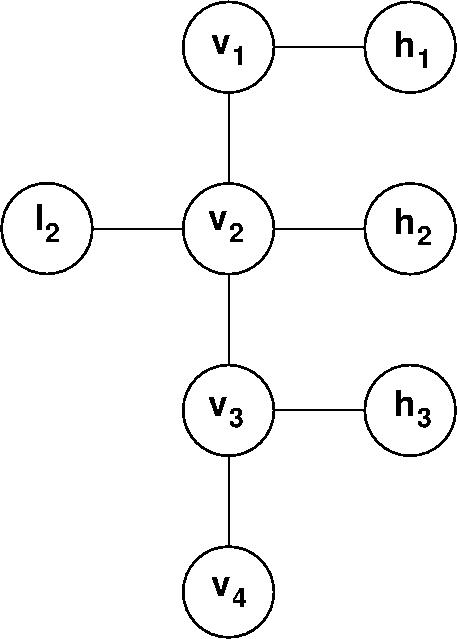}
\caption{An example for a caterpillar graph}
\end{figure}

In chemical graph theory, caterpillar graphs are useful in studying 
topological properties of benzenoid hydrocarbons[2]. In fact, Basil and
Sherif [2]observe that\\

``It is amazing that nearly all graphs that played an important role in
what is now called ``chemical graph theory'' may be related to
caterpillar trees.'' \\

Ortiz and Villanueva [4] describe an algorithm for enumerating a family
of maximal independent sets in caterpillar graphs. The algorithm takes
time linear in the size of the output, i.e., is linear in the sum of
sizes of all maximal independent sets.  They also propose $CC^2(P_k)$, a
generalisation in which hairs have length exactly two. 

In this paper, we consider a still more generalised version of
caterpillar graphs (See Figure 2).  In this generalised version, we allow
a backbone vertex $v_i$ to have up to one hair of length two and any
number of hairs of length one; in particular $v_i$ may have no hairs at
all.  If $P_k=\{v_1,... ,v_k\}$, a chordless path of length $k$ is the
backbone, then we denote the generalised caterpillar graph by
$C^{1,2}(P_k)$.

\begin{figure}[h]
\centering 
\includegraphics[width=2in]{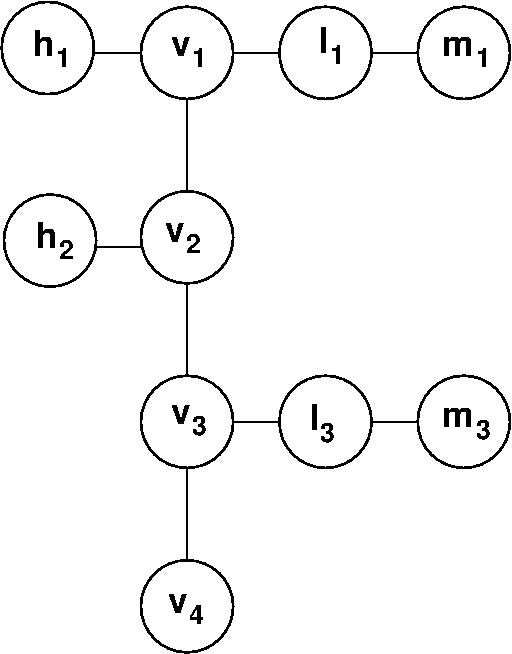} 
\caption{An example for a generalised caterpillar graph}
\end{figure} 

A complete caterpillar graph $CC(P_k)$ is a (usual) caterpillar graph
such that there is at least one hair at each of its backbone vertices.
The graph of Figure 1 is not complete as vertex $v_4$ does not have any
hair. The contraction graph $G_k$ of a (usual) caterpillar graph $C(P_k)$
is the graph obtained by contracting, for each backbone vertex $v_i$ of
the $C(P_k)$, all the pendant vertices incident at $v_i$ to a single
vertex [4].

An independent set or a stable set is a set of vertices in a graph such
that no two vertices in the set are adjacent. That is, it is a set $I$ of
vertices of a graph $G$ such that if $I$ contains two vertices, say $a$
and $b$, then $ab$ is not an edge of $G$. The size or cardinality of an
independent set $I$ is the number of vertices in the set $I$. An
independent set $I$ will be called a maximal independent set if every
vertex $v$ is either in $I$ or is adjacent to a vertex in $I$.

Valiant [6] shows that the problem of counting the number of maximal
independent sets is $\# P$-complete for general graphs. Tsukiyama et al.
[5] show that we can enumerate a family of maximal independent sets of a
general connected graph in $O(n m m(G))$ time; here $n$ is the number of
vertices, $m$ is the number of edges and $m(G)$ is the number of maximal
independent sets of a graph $G$. Ortiz and Villanueva [4] show that
$m(C(P_k))$, the number of maximal independent sets of a (usual)
caterpillar graph $C(P_k)$ is the same as $m(G_k)$, the number of maximal
independent sets of its contraction graph, $G_k$.  They give an algorithm
to find a family of maximal independent sets of a caterpillar graph in
time polynomial in the number of maximal independent sets. In this paper,
we obtain a similar result for generalised caterpillar graphs. 

Vertices of generalised caterpillar graph can be partitioned into stages.
Vertex $v_i$ together with vertices in hairs incident at $v_i$ will be
said to be at stage $i$. Formally, if we delete vertex $v_i$ from
$C^{1,2}(P_k)$, the graph may split into several components. The vertices
in components which do not contain either vertex $v_{i-1}$ or $v_{i+1}$
will be at stage $i$ (along with vertex $v_i$). 

Let $x_1,x_2, ..., x_{m_i}$ be the pendant vertices in hairs of length
one at stage $i$ of $C^{1,2}(P_k)$. If any of these vertices is in a
maximal independent set $S$, then $v_i$ cannot be in the independent set
$S$. Conversely, if $v_i$ is not in $S$, then we have to put all these
vertices in the maximal independent set $S$. Thus, any pendant vertex
belonging to a hair of length one at stage $i$ is contained in a maximal
independent set $S$, if and only if all pendant vertices at stage $i$ are
contained in $S$.

Hence, the number of maximal independent sets of $C^{1,2}(P_k)$ is
independent of the number of hairs of length one at each stage (provided
there is at least one such hair).  Hence from now on, we assume that
$C^{1,2}(P_k)$ has at most one hair of length one at each stage $i$
(wherever there is at least one such hair) and we denote this hair by
$v_ih_i$.  We will denote the hair of length two at stage $i$ by
$v_il_im_i$, wherever it exists.

\section{Structure of Maximal Independent Sets in generalised Caterpillar
Graphs}

If $S$ is any maximal independent set of $C^{1,2}(P_k)$, let $S_i$ be the
subset of $S$ containing only vertices at stage $i$. Clearly, $S=S_1 \cup
S_2 \cup ...  S_k$.  In a caterpillar graph, the only vertices from two
different stages, adjacent to each other are the vertices $v_i$ and
$v_{i+1}$ (or $v_{i-1}$ and $v_i$).

In case, if a hair of length one is present at stage $i$ (hair of length
two may or may not be there), then either $h_i$ or $v_i$ (but not both)
will be present in any maximal independent set. Hence, in this case,
exactly one of $h_i$ or $v_i$ will be in the set $S_i$.

Similarly, if a hair of length two is present at stage $i$ (hair of
length one may or may not be present), then either $m_i$ or $l_i$ (but
not both) will be present in any maximal independent set. Hence, in this
case, exactly one of $m_i$ or $l_i$ will be present in $S_i$.

Also observe that as $v_i$ and $l_i$ are adjacent, both of them cannot be
present in any independent set.

In general, either both hairs of length one and two, or only one or
neither may be present at stage $i$. In all there are exactly three
possibilities. \\

\noindent {\textbf{Case 1 ({$v_i\in S_i$}):}} If $v_i\in S_i$,
then if a hair of length two is present then $l_i\not\in S_i$ and hence
$m_i\in S_i$. Moreover, even if we have hair of length one, then as $v_i$
and $h_i$ are adjacent, $h_i\not\in S_i$, thus \\

\noindent \noindent $S_i= \{v_i,m_i\}$ if hair of length two is present \\ $S_i= \{v_i\}$
otherwise\\

\noindent {\textbf{Case 2 ({$l_i\in S_i$}):}} If $l_i\in S_i$, then $v_i\not\in
S_i$ and $m_i\not\in S_i$. If a hair of length one is present at stage
$i$, then as $v_i\not\in S_i$, $h_i\in S_i$, thus, in this case

\noindent $S_i= \{h_i,l_i\}$ if hair of length one is present \\ $S_i=\{l_i\}$ 
otherwise\\

\noindent {\textbf{Case 3 ({$v_i,l_i\not\in S_i$})}} If length one hair is present,
then as $v_i\not\in S_i$ we must have $h_i\in S_i$. If length two hair is
present then as $l_i\not\in S_i$, $m_i\in S_i$. Hence, in this case,

\noindent $S_i= \{h_i,m_i\}$  if hairs of length one
and two are both present \\ $S_i=\{h_i\}$  if only hair of length one
is present \\ $S_i=\{m_i\}$  if only hair of length two is present \\
$S_i=\Phi$ otherwise\\

Thus, in each of these cases, all vertices at stage $i$, except possibly
$v_i$, will either be in the independent set $S_i$ or will be adjacent to
a vertex in $S_i$. If $v_i\in S_i$, the set $S_i$ will be a maximal
independent set of the subgraph at stage $i$.

Thus, we have the following lemma:

{\textbf{Lemma 1:}} $S=S_1 \cup S_2 \cup ...  S_k$ is a Maximal
Independent Set of generalised caterpillar graph $C^{1,2}(P_k)$ if and
only if following conditions hold: \\ (1) for each $v_i$ which is not
adjacent to a vertex in $S_i$, either $v_{i-1}\in S_{i-1}$ or $v_{i+1}\in
S_{i+1}$.\\ (2) both $v_{i-1}\in S_{i-1}$ and $v_i\in S_i$ should not
simultaneously hold.

{\textbf{Proof:}} As each $S_i$ is an independent set, and as each vertex
of $S_i$ except possibly $v_i$ is either in $S_i$, or has a neighbour in
$S_i$, the set $S$ will be a maximal independent set iff either each
$v_i\in S$ or each $v_i$ has a neighbour in $S$.  Vertex $v_i\in S$ iff
$v_i\in S_i$.

If $v_i\not\in S_i$, then $v_i$ has a neighbour in $S$ iff one of the
following conditions hold\\ (1) either $v_i$ has a neighbour in $S_i$,
or\\ (2) $v_{i-1}\in S$ or\\ (3) $v_{i+1} \in S$.

For the set $S$ to be independent, clearly the second condition must
hold. The lemma thus follows. []

\section{Finding Maximal Independent Sets in generalised Caterpillar
Graphs}

Let us assume that $S=S_1 \cup S_2 \cup ...  S_k$ is a maximal
independent set of $C^{1,2}(P_k)$.  
Then, for $i \leq k$, we classify the set $S_i$ depending upon the
``status'' of vertex $v_i$.\\

\noindent {\textbf{Type 1:}} If $v_i \in S_i$, then we will say $S_i$ is of Type 1.

\noindent {\textbf{Type 2:}} If $v_i\not\in S_i$, but some neighbour of $v_i$ is in
$S_i$, then we will say $S_i$ is of Type 2. In this case either $h_i \in
S_i$ or $l_i \in S_i$.

\noindent {\textbf{Type 3:}} If no neighbour of $v_i$ is in $S_i$, but $v_{i-1}\in
S_{i-1}$, then we will say $S_i$ is of Type 3. 

\noindent {\textbf{Type 4:}} If $v_{i-1}\not\in S_{i-1}$, $v_i\not\in S_i$ and no
neighbour of $v_i$ is in $S_i$, then we will say that $S_i$ is of Type 4.
In this case, for the set $S$ to be maximal, $v_{i+1}\in S_{i+1}$.

The last stage $S_k$ cannot be of Type 4, as in that case neither $v_k\in
S$ nor $v_{k-1}\in S$. And hence, as $v_k$ does not have a neighbour in
$S_k$, $v_k$ will not have any neighbour in $S$ (there is no vertex
$v_{k+1}$), violating maximality. Further, as $S_1$ is the first
independent set, $S_1$ cannot be of Type 3 (there is no vertex $v_0$).

We store the types of hairs present at stage $i$ in an array $T$.  The
entry \\
$T[i]= 0$  if there are no hairs at stage  $i$\\
$T[i]=	1$ if there are only length one hairs at stage $i$\\
$T[i]=	2$ if there is only length two hair at stage  $i$\\
$T[i]=	3$ if stage  $i$  has both types of hairs \\

Then the table below summarises the discussion above and gives the list
of all $S_i$'s of each type.
\begin{table}[h!]
\centering
\begin{tabular}{|c||c|c|c|c|}
  \hline
  $T[i]$ 	& $S_i$ of type 1 	& $S_i$  of type 2  &$S_i$ of type 3 &$S_i$ of type 4\\
  \hline \hline	
  0  	 	& $\{v_i\}$ 		&none			  	& $\phi$  		& $\phi$  \\
  \hline
  1  	 	& $\{v_i\}$ 		& $\{h_i\}$		  	& none 			& none  \\
  \hline
  2  	 	& $\{v_i,m_i\}$ 	& $\{l_i\}$		  	& $\{m_i\}$  		& $\{m_i\}$  \\
  \hline
  3  	 	& $\{v_i,m_i\}$ 	& $\{h_i,l_i\}$,$\{h_i,m_i\}$  	& none  		& none  \\
  \hline
\end{tabular}
\caption{The possible instances of $S_i$ and their types depending on the hairs present}
\end{table}

{\textbf{Theorem 1:}} $S= S_1 \cup S_2 \cup ...  S_k$ is a maximal
independent set of a generalised caterpillar graph $C^{1,2}(P_k)$ if and
only if for $1\leq i \leq k-1$, $(S_i,S_{i+1})$ is of one of the
following forms, and $S_k$ is not of the type 4.\\ (1) (type 1, type 2)\\
(2) (type 1, type 3)\\ (3) (type 2, type $x$), where $x$ is 1,2 or 4\\
(4) (type 3, type $x$), where $x$ is 1,2 or 4\\ (5) (type 4, type 1)

{\textbf{Proof:}} Let us prove the `only if' part first. Let us assume
that $S$ is a maximal independent set of $C^{1,2}(P_k)$. We have one of
the following three cases, depending on the type of $S_i$.\\
(1)  If $S_i$ is of Type 1, then as $v_i\in S_i$, $S_{i+1}$ cannot be
of Type 4. As $v_i\in S_i$, $v_{i+1} \notin S_{i+1}$, hence $S_{i+1}$
cannot be of Type 1. If $v_{i+1}$ has a neighbour in $S_{i+1}$, then
$S_{i+1}$ will be of Type 2, otherwise of Type 3.

\noindent (2) If $S_i$ is of Type 4, then as we saw earlier, $v_{i+1}\in S_{i+1}$,
and hence $S_{i+1}$ will be of Type 1.

\noindent (3) If $S_i$ is of Type 2 or of Type 3, then as $v_i \notin S_i$,
$S_{i+1}$ cannot be of Type 3. If $v_{i+1}\in S_{i+1}$, then $S_{i+1}$
will be of Type 1. If a neighbour of $v_{i+1}$ is in $S_{i+1}$, then
$S_{i+1}$ will be of Type 2. If $v_{i+1}\not\in S_{i+1}$, and if it does
not have any neighbour in $S_{i+1}$, then $S_{i+1}$ will be of Type 4.

To prove the `if' part, we need to prove that for $1\leq i \leq k$, if
$(S_i,S_{i+1})$ is in one of the forms listed (and $S_k$ is not of Type
4), then $S$ is a maximal independent set. As each vertex at stage $i$,
except possibly $v_i$, is either in $S_i$ or is adjacent to a vertex in
$S_i$, we only need to show that \\ (a) Both $v_i\in S_i$ and $v_{i+1}\in
S_{i+1}$ cannot simultaneously hold.\\ (b) If $v_i\not\in S_i$, then
$v_i$ has a neighbour in $S$.

If $v_i\in S_i$, then $S_i$ will be of Type 1, and if $v_{i+1}\in
S_{i+1}$, then $S_{i+1}$ will also be of Type 1. As we do not have the
form (Type 1,Type 1), the first condition holds.

If $v_i\not\in S_i$, then $S_i$ will not be of Type 1. If $S_i$ is of
Type 2, then $v_i$ will have a neighbour in $S_i$, and hence in $S$. If
$S_i$ is of Type 3, then as the only permissible form for $(S_{i-1},S_i)$
is (Type 1,Type 3). Thus, $S_{i-1}$ has to be of Type 1, and $v_{i-1}\in
S_{i-1}$; or $v_i$ will have a neighbour $v_{i-1}$ in $S_{i-1}$, and
hence in $S$.

Finally, if $S_i$ is of Type 4, then as the only permissible form is
(Type 4, Type 1). Thus, $S_{i+1}$ has to be of Type 1 and $v_{i+1}\in
S_{i+1}$ (for $i\neq k$) and so $v_{i+1}$ will have a neighbour in
$S_{i+1}$, or in $S$. Hence, $S$ will be both independent and maximal. []

\section{Finding  Family of Maximal Independent Sets}

From Theorem 1, if $(S_i,S_{i+1})$ is of one of the forms listed, then
$S=S_1\cup S_2 \cup ... S_k$ will be a maximal independent set. Thus, to
find a family of all maximal independent sets, we need to find all such
valid sequences. For this, we construct a directed $k$-level graph $L_k$
such that any maximal independent set in the generalised caterpillar
graph corresponds to a source-sink (source to sink) path in $L_k$. 

A $k$-level graph $G=(V,E,\phi)$ with $k \leq n$ is a graph with an
assignment of levels $\phi:V \rightarrow \{1,2, ..., k\}$ that
partitions the vertex set into $k$ pairwise disjoint subsets, $V_1, V_2,
..., V_k$ such that $V=V_1 \cup V_2 \cup ... V_k$. Further, if $(uv)$ is
an edge in $G$, then $u$ and $v$ are not in the same level [1]. In our
level graph $L_k$, the edges are only from level $i$ to level $i+1$.

We will denote the set of vertices of $L_k$ by $U$.  The $k$ levels in
$L_k$ are numbered from 1 to $k$; level $i$ in $L_k$ corresponding to
stage $i$ in $C^{1,2}(P_k)$.  Roughly speaking, at each level $i$, the
vertices in $L_k$ will correspond to one possible instance of $S_i$.  The
total number of vertices present at any level $i$ of the level graph,
will depend on the types of hairs present at stage $i$. We will see that 
the number of vertices at each level will be at most five.

We will use two labels ``$type$'' and ``$index$'' on vertices of $L_k$.
Type of a (new) vertex will correspond to the type of corresponding
$S_i$. Index will be one in all but one case. From Table 1, observe that,
in all but one case, for any value of $T[i]$, there is at most one
instance of $S_i$. If $T[i]=3$, then there are two instances of $S_i$. We
use the label ``$index$'' to distinguish these cases.

In more detail, we add vertices at level $i$ as follows. First, for each
$i$, we first add a vertex $p_i$ and set $type(p_i)=1$ and
$index(p_i)=1$. This will correspond to the case when $v_i\in S_i$.

Depending upon type of hairs present at $v_i$, we add other vertices at
level $i$ as follows:

\noindent {\textbf{Only Length one hairs ($T[i]==1$):}} Add a vertex $s_i$ and set
$type(s_i)=2$ and $index(s_i)=1$. This will correspond to the case when
$S_i$ is of type 2, and $S_i=\{h_i\}$.

\noindent {\textbf{Only Length two hairs ($T[i]==2$):}} In this case, also, we
first add a vertex $s_i$ and set $type(s_i)=2$ and $index(s_i)=1$. This
will correspond to the case when $S_i$ is of type 2, and $S_i=\{l_i\}$.  
Further, we also 

\noindent (a) for $i\geq 2$, we add a vertex $t_i$ and set $type(t_i)=3$ and
$index(t_i)=1$.
  
\noindent (b) and for $i\leq k-1$, we add a vertex $u_i$ and set $type(u_i)=4$ and
$index(u_i)=1$.  

These cases correspond to the case when $S_i$ is of type 3 or of type 4.
In these cases, $S_i=\{m_i\}$.

\noindent {\textbf{Both Length one and two hairs ($T[i]==3$)}}: Add two vertices
$q_{i}$ and $r_{i}$ and set $type(q_{i})=type(r_{i})=2$; $index(q_{i})=1$
and $index(r_{i})=2$.

This corresponds to the case when $S_i$ is of type 2. Here $S_i$ can be
either $\{h_i,l_i\}$ or $\{h_i,m_i\}$, thus we need two vertices, one for
each case. We also use the label ``$index$'' to distinguish these two
cases.

\noindent {\textbf{{No Hairs ($T[i]==0$):}}} In this case

\noindent (a) for $i\geq 2$, we add a vertex $t_i$ and set $type(t_i)=3$ and
$index(t_i)=1$.
  
\noindent (b) and for $i\leq k-1$, we add a vertex $u_i$ and set $type(u_i)=4$ and
$index(u_i)=1$.  

These cases correspond to the case when $S_i$ is of type 3 or of type 4.
In these cases, $S_i=\Phi$.

This is summarised in Table 2 below:
\begin{table}[h!]
\centering
\begin{tabular}{|c||c|c|c|l|}
  \hline
  vertex & type    &index  & Value of $T[i]$ & set $S_i$ \\
  \hline \hline	
  $p_i$  & 1	   & 1	   & $0,1,2$ or $3$ & $v_i\in S_i$ \\
  \hline
  $s_i$  & 2 	   & 1	   & $1$ or $2$ & $S_i=\{h_i\}$ or $S_i=\{l_i\}$ \\
  \hline
  $q_i$  & 2	   & 1	   & $3$ & $S_i=\{h_i,l_i\}$ \\
  \hline
  $r_i$  & 2	   & 2	   & $3$ & $S_i=\{h_i,m_i\}$ \\
  \hline
  $t_i$  & 3	   & 1	   & $0$ or $2$ & $S_i=\{m_i\}$ or $S_i=\Phi$ \\
  \hline
  $u_i$  & 4	   & 1	   & $0$ or $2$ & $S_i=\{m_i\}$ or $S_i=\Phi$ \\
  \hline
\end{tabular}
\caption{The vertices at level $i$ 
in $L_k$ and the corresponding $S_i$}
\label{tab:sec}
\end{table}

Next we add following edges. We add an edge from a vertex $a$ in level
$i$ to a vertex $b$ in the next level $i+1$, if $(type(a),type(b))$ is
one of the following (listed in Theorem 1):\\

$(1,2)$, $(1,3)$, $(2,1)$, $(2,2)$, $(2,4)$, $(3,1)$, $(3,2)$, $(3,4)$ or
$(4,1)$

Formally, we add the following edges from a vertex $a$ of level $i$ to a
vertex $b$ of level $i+1$, for $1\leq i\leq k-1$:

\noindent (a) For each vertex $a$ of type 1, we add an edge $(a,b)$ in $L_k$, if
vertex $b$ is either of type 2 or of type 3.
 
\noindent (b) For each vertex $a$ of type 2 or type 3, we add an edge $(a,b)$, if 
vertex $b$ is of type 1, type 2, or of type 4 (i.e., $b$ is not of type
3).

\noindent (c) For each vertex $a$ of type 4 we add an edge $(a,b)$, if vertex $b$
is of type 1.

All vertices at Level 1 will be treated as sources and all vertices at
level $k$ as sinks, and hence any path from level $1$ to level $k$ in
$L_k$ will be a source-sink path.

As these edges correspond exactly to valid $(S_i,S_{i+1})$ pairs of
Theorem 1, hence again from Theorem 1, it follows that any source-sink
path in $L_k$ will correspond to a valid set of sequence $S_1, S_2, ...,
S_k$ and conversely.

Thus, if we find all source-sink paths in $L_k$, we can find all the
maximal independent sets in $C^{1,2} (P_k)$. This can be done by using a
generalised depth first procedure (DFS), similar to the one used by [4].

We basically, put the start vertex in a stack and then for each neighbour
of the start vertex (as the new start vertex), we call the procedure
recursively. We stop when the start vertex for the current call is a sink
vertex, and print the entire stack; and also remove this vertex from the
stack (backtrack to the previous level). 

The number of source-sink paths in $L_k$ will be the number of maximal
independent sets in $C^{1,2}(P_k)$.  We can easily obtain maximal
independent sets from the source-sink path.  Let $w_1,w_2, ..., w_k$ be
the vertices in a source sink path of $L_k$. Then we reconstruct the
$S_i$ corresponding to each $P_i$ as follows (see Table 1):\\

\noindent If $type(w_i)==1$, then \\ if $T[i]==3$ or $T[i]==2$ then
$S_i=\{v_i,m_i\}$\\ else $S_i=\{v_i\}$\\

\noindent If $type(w_i)==3$ or $type(w_i)==4$, then\\ if $T[i]==2$ then
$S_i=\{m_i\}$\\ if $T[i]==0$ then $S_i=\Phi$\\

\noindent If $type(w_i)==2$, then \\ if ($T[i]==1$) then $S_i=\{h_i\}$\\ if
($T[i]==2$) then $S_i=\{l_i\}$\\ if $T[i]=3$\\ \indent if $index(w_i)==1$
then $S_i=\{h_i,l_i\}$,\\ \indent else if $index(w_i)==2$ then
$S_i=\{h_i,m_i\}$\\

A theorem similar to that of Ortiz and Villanueva [4], also holds for the
generalised caterpillar graph.

{\textbf{Theorem 2:}} We can enumerate all maximal independent sets of
$C^{1,2}(P_k)$ in $O(k m(C^{1,2}(P_k)))$ time, where $m(C^{1,2}(P_k))$ is
the number of maximal independent sets of $C^{1,2}(P_k)$.

{\textbf{Proof:}} For each source-sink path $P$ of $L_k$, generalised
depth first procedure is called once for each vertex of $P$. Since any
edge in $L_k$ is between adjacent levels, the number of vertices in the
path is $k+1$. Hence there are only $O(k)$ calls to the procedure.  As
each call takes $O(1)$ time, except when we reached a sink, in which case
it takes $O(k)$ time. But as we reach sink only once for a path, the
algorithm takes $O(k)$ time, for each path found. 

As there are $m(C^{1,2}(P_k))$ such paths, the algorithm takes $O(k 
m(C^{1,2}(P_k)) )$ time. []

As each maximal independent set has $k$ vertices, the algorithm is
essentially linear in the output size.

\section{Example}

Let us illustrate the algorithm for the graph given in Figure 2, we have
the following vertices of type 1 in it:

\noindent Level 1: $p_1$, $type(p_1)=1$, $index(p_1)=1$
 
\noindent Level 2: $p_2$, $type(p_2)=1$, $index(p_2)=1$
 
\noindent Level 3: $p_3$, $type(p_3)=1$, $index(p_3)=1$
 
\noindent Level 4: $p_4$, $type(p_4)=1$, $index(p_4)=1$

We have the following vertices of type 2: 

\noindent Level 1: $T[i]=3$.  Two vertices $q_1$ and $r_1$, $type(q_1)=2$,
$index(q_1)=1$, $type(r_1)=2$, $index(r_1)=2$
 
\noindent Level 2: $T[i]=1$. Vertex $s_2$, $type(s_2)=2$, $index(s_2)=1$
 
\noindent Level 3: $T[i]=2$. Vertex $s_3$, $type(s_3)=2$, $index(s_3)=1$
 
\noindent Level 4: $T[i]=0$. No vertex.

We have the following vertices of type 3 in it: 

\noindent Level 1: $T[i]=3$.  None 
 
\noindent Level 2: $T[i]=1$.  None
 
\noindent Level 3: $T[i]=2$.  Vertex $t_{3}$, $type(t_3)=3$, $index(t_3)=1$
 
\noindent Level 4: $T[i]=0$.  Vertex $t_{4}$, $type(t_4)=3$, $index(t_4)=1$

The following are vertices of type 4.

\noindent Level 1: $T[i]=3$.  None
 
\noindent Level 2: $T[i]=1$.  None
 
\noindent Level 3: $T[i]=2$. Vertex $u_3$, $type(u_3)=4$, $index(u_3)=1$
 
\noindent Level 4: $T[i]=0$. None as this is the last level

The edges in the example are as shown in Figure 3.

\begin{figure}[h]
 \centering
   \includegraphics[width=4in]{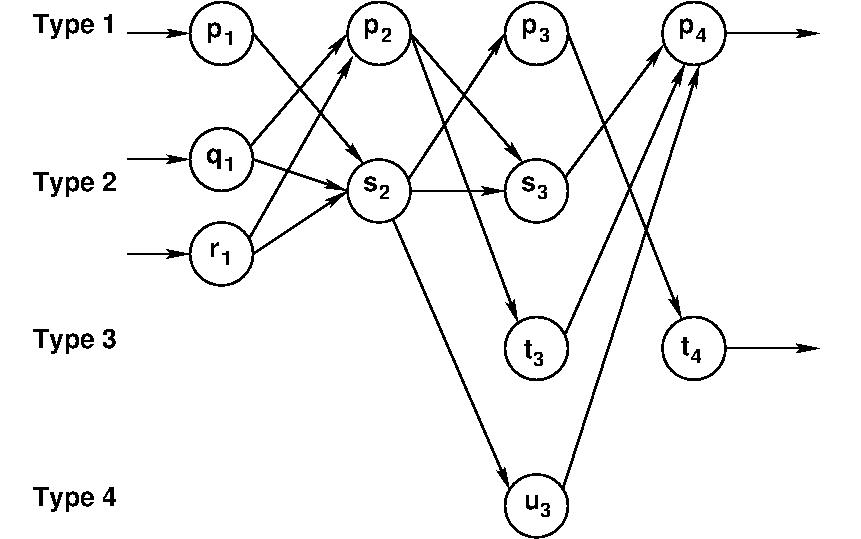}
  \caption{Construction of $L_k$ from $C^{1,2}(P_k)$}
\end{figure}

Each source-sink path corresponds to a maximal independent set. For
example, the path $p_1-s_2-u_3-p_4$ in $L_k$ corresponds to the maximal
independent set $\{v_1,m_1,h_2,m_3,v_4\}$ in $C^{1,2}(P_k)$. Similarly,
the path $r_{1}-s_2-p_3-t_4$ in $L_k$ corresponds to the maximal
independent set $\{h_1,m_1,h_2,v_3,m_3\}$ in $C^{1,2}(P_k)$.

\section*{Conclusions}

We discuss the problem of finding a family of maximal independent sets in
a generalised caterpillar graph. We show that this problem can also be
reduced to the problem of finding all source-sink paths in a level graph.
The proposed algorithm takes time linear in the output size (total number
of vertices in all maximal independent sets).

Further, we believe, that this algorithm can be extended for another
generalisation of caterpillar graph, where each vertex of backbone has
bounded number of hairs of length more than one. It may also be possible
to generalise the algorithm for some other generalisations of caterpillar
graphs, possibly having a different set of hairs.  We may have to
identify the new set of possible $S_i$'s at each stage $i$ and
classifying them into possibly some other different ``types''. 

\section*{References}

\noindent [1] Christian Bachmaier and Franz J. Brandenburg. {\em Circle Planarity
of Level Graphs}. PhD thesis, Faculty of Mathematics and Computer
Science, University of Passau, 2004.

\noindent [2] El-Basil and Sherif. Applications of caterpillar trees in chemistry
and physics. {\em Journal of Mathematical Chemistry}, 1:153--174, 1987.

\noindent [3] F. Harary and A. J. Schwenk. The number of caterpillars.  {\em
Discrete Mathematics}, 6:359--365, 1973.

\noindent [4] Carmen Ortiz and Monica Villanueva. Maximal independent sets in
caterpillar graphs. {\em Discrete Appl. Math.}, 160(3):259--266, 2012.

\noindent [5] Shuji Tsukiyama, Mikio Ide, Hiromu Ariyoshi, and Isao Shirakawa. A
new algorithm for generating all the maximal independent sets. {\em SIAM
J. Comput.}, 6(3):505--517, 1977.

\noindent [6] Leslie G. Valiant. The complexity of computing the permanent. {\em
Theor. Comput. Sci.}, 8:189--201, 1979.

\end{document}